\documentclass{pasj00}

\usepackage{color}
\newcommand{\foxsi}[0]{\textit{FOXSI }}
\newcommand{\rhessi}[0]{\textit{RHESSI }}
\newcommand{\hinode}[0]{\textit{Hinode }}

\begin{document}
\SetRunningHead{Ishikawa et al.}{Constraining hot plasma with FOXSI}

\title{Constraining Hot Plasma in a Non-flaring Solar Active Region with FOXSI Hard X-ray Observations}

\author{Shin-nosuke \textsc{Ishikawa} %
}
\affil{National Astronomical Observatory of Japan, 2-21-1 Mitaka, Tokyo 181-8588, Japan}
\email{s.ishikawa@nao.ac.jp}
\author{Lindsay \textsc{Glesener}}
\affil{Space Science Laboratory, University of California, Berkeley, CA 94720, USA}\email{glesener@ssl.berkeley.edu}
\author{Steven \textsc{Christe}}
\affil{NASA Goddard Space Flight Center, Greenbelt, MD 20771-0001, USA}\email{steven.d.christe@nasa.gov}
\author{Kazunori \textsc{Ishibashi}}
\affil{Nagoya University, Furo-cho, Chikusa-ku, Nagoya, Aichi 464-8602, Japan}\email{bish@u.phys.nagoya-u.ac.jp}
\author{David H. \textsc{Brooks}}
\affil{ College of Science, George Mason University, 4400 University Drive, Fairfax, VA 22030, USA}\email{dhbrooks@ssd5.nrl.navy.mil}
\author{David R. \textsc{Williams}}
\affil{Mullard Space Science Laboratory, University College London, Holmbury St Mary, Surrey, RH5 6NT, UK}\email{d.r.williams@ucl.ac.uk}
\author{Masumi \textsc{Shimojo}}
\affil{National Astronomical Observatory of Japan, 2-21-1 Mitaka, Tokyo 181-8588, Japan}\email{masumi.shimojo@nao.ac.jp}
\author{Nobuharu \textsc{Sako}}
\affil{National Astronomical Observatory of Japan, 2-21-1 Mitaka, Tokyo 181-8588, Japan}\email{sako@solar.mtk.nao.ac.jp}

\and
\author{S\" am \textsc{Krucker}}
\affil{Space Science Laboratory, University of California, Berkeley, CA 94720, USA, \\
and Institute of 4D Technologies, School of Engineering, University of Applied Sciences and Arts Northwestern Switzerland, 5210 Windisch, Switzerland}\email{krucker@ssl.berkeley.edu}

%


\maketitle

\begin{abstract}
We present new constraints on the high-temperature emission measure of a non-flaring solar active region using observations from the recently flown \textit{Focusing Optics X-ray Solar Imager} sounding rocket payload.  \foxsi has performed the first focused hard X-ray (HXR) observation of the Sun in its first successful flight on 2012 November 2. Focusing optics, combined with small strip detectors, enable high-sensitivity observations with respect to previous indirect imagers. This capability, along with the sensitivity of the HXR regime to high-temperature emission, offers the potential to better characterize high-temperature plasma in the corona as predicted by nanoflare heating models.  We present a joint analysis of the differential emission measure (DEM) of active region 11602 using coordinated observations by \textit{FOXSI}, \textit{Hinode}/XRT and \textit{Hinode}/EIS.  
The \textit{Hinode}-derived DEM predicts significant emission measure between 1 MK and 3 MK, with a peak in the DEM predicted at 2.0--2.5 MK.  
The combined XRT and EIS DEM also shows emission from a smaller population of plasma above 8 MK. 
This is contradicted by \foxsi observations that significantly constrain emission above 8 MK.  
This suggests that the \textit{Hinode} DEM analysis has larger uncertainties at higher temperatures 
and that $>$8 MK plasma above an emission measure of 3$\times$10$^{44}$~cm$^{-3}$ is excluded in this active region.  
\end{abstract}

\section{Introduction}

The energy source of the high-temperature corona is one of the main open questions in heliophysics. The two leading theoretical solutions are (1) heating by a large number of small-scale magnetic energy releases in the chromosphere and corona (so-called `nanoflare heating') or (2) heating by waves that are excited below the photosphere 
and released in the chromosphere and corona (`wave heating'). One of the main differences between these two models is the prediction of the time evolution of the plasma temperature as the heat input and output (e.g. radiative and conductive transfers) interact (e.g. \cite{cargill2004, vanballegooijen2011}). This difference is most easily discerned by measuring the amount of plasma at each temperature through the differential emission measure (DEM) distribution from below one million up to $\sim$20 million Kelvin.

The most accurate DEM measurements are from non-flaring active regions (e.g. \cite{reale2009a, patsourakos2009, odwyer2011, testa2011, warren2011, warren2012, warren2013, hannah2012})
where non-equilibrium effects are potentially less important than during flares (e.g. \cite{reale2008}). Observations in the extreme ultra-violet (EUV), soft X-rays (SXRs), and hard X-rays (HXRs) provide constraints of the DEM in different temperature ranges. To get information over a wide temperature range, it is essential to combine multiwavelength observations. 

Currently-available observations in the EUV by \hinode and the \textit{Solar Dynamics Observatory (SDO)} provide accurate information about the low temperature plasma ($<$8 MK) but emission at the high-temperatures ($>$8 MK) is generally too weak to be detected by currently available EUV line observations.  
Past observations from \textit{Hinode}'s EUV Imaging Spectrometer (EIS) have characterized the DEM in the few million Kelvin range (e.g. \cite{warren2012}) and have found a DEM for non-flaring active regions with a peak around $\sim$4 MK. The observed slope of the DEM is rather steep, favoring wave heating models. EUV and SXR broadband filter observations cover a wider range in temperature, but observations are not without intrinsic biases (e.g. \cite{schmelz2009b, winebarger2012, guennou2013}).  
Besides the few-MK component in non-flaring active regions, EUV and SXR filter observations suggest the existence of an additional, hotter component that is speculated to be a direct signature of nanoflare heating (e.g. \cite{reale2009a, schmelz2009a, testa2012}), 
although not without controversy (e.g. \cite{schmelz2009b, winebarger2012}). Others report no significant component above the sensitivity limit of these instruments (e.g. \cite{testa2011}). In any case, current observations show that an additional hot component, if present, is much weaker than the main emission around a few MK, making an unambiguous detection difficult. Observations of a high temperature component is limited as currently-available EUV and SXR instrumentation have `blind spots' at higher temperatures \citep{winebarger2012}, particularly for non-flaring observations that have an intrinsically low high-temperature emission measure. Future observations of high-temperature SXR lines have the potential to significantly improve high-temperature diagnostics, and appropriate instrumentation is under development \citep{kobayashi2011}. 

To restrict high-temperature ($>$8 MK) plasma, X-ray observations of the thermal bremsstrahlung emission of hot electrons provide excellent sensitivity (e.g. \cite{mctiernan2009, reale2009b, miceli2012}).  
By combining EUV and SXR diagnostics with observations in HXRs from the \textit{Reuven Ramaty High Energy Spectroscopic Imager} (\textit{RHESSI}, \cite{lin2002}), the high-temperature component in an active region DEM is sometimes significantly restricted, indicating that EUV and SXR alone can overestimate hot emissions \citep{schmelz2009b}, in particular in the presence of strong emissions at lower temperatures. Using X-ray observations only, \citet{mctiernan2009} showed that there are indeed non-flaring times (defined as times without a \textit{GOES} flare) with a significant 5-10 MK component in the DEM. Since the appearance of this component scales with the solar cycle, it is, however, unclear whether the observed component is due to unresolved microflares not excluded in the selection of non-flaring times or due to continuous heating relevant to overall coronal heating. Unfortunately, \rhessi observations  provide only limited sensitivity due to the high nonsolar background of its bulk germanium detectors.  

In this paper, we present an active region measured in HXRs with greatly enhanced sensitivity using focusing optics on the recently flown \textit{Focusing Optics X-ray Solar Imager (FOXSI)} sounding rocket payload.  \foxsi is funded under NASA's Low Cost Access to Space (LCAS) program and is the first example of HXR focusing optics applied to solar observation.  
\foxsi is a HXR imaging spectrometer operating in the energy range of 4--15 keV, with an angular resolution of $\sim$9 arcseconds (on-axis) and an energy resolution of $\sim$0.5 keV \citep{krucker2013foxsi}.

\foxsi flew for the first time, successfully, on 2012 November 2 from the White Sands Missile Range in New Mexico, USA \citep{krucker2014}. Targets included active regions on the disk and quiet portions of the Sun.  A microflare (\textit{GOES} class B2.7) was observed by \foxsi on the western limb \citep{krucker2013foxsi}, serving as useful in-flight calibration of the instrument, since it was also observed by \textit{RHESSI}.  In this paper we report on the \foxsi observation of active region 11602 in the $>$4 keV range and compare to EUV and SXR observations from the \hinode \citep{kosugi2007} EUV Imaging Spectrometer (EIS, \cite{culhane2007})  and X-ray Telescope (XRT, \cite{golub2007}) to constrain a potential hot component.

\section{Observations}
\foxsi observed the Sun beginning at 2012 November 2 17:56:48 UT for a total of 6.5 minutes, with 44 seconds of observation time on active region (AR) 11602 (its first target). Observations were coordinated with the \hinode spacecraft (\hinode Operation Plan \#221). 
\hinode observed AR 11602 before and after the \foxsi launch, from 16:44 to 17:50~UT and from 19:00 to 20:15~UT.  The XRT and EIS instruments ran programs appropriate for the measurement of active region DEMs with multiple XRT filter pairs and multiple EIS wavelength bands to cover the temperature range of log~$T \sim$5.5 -- 7.2.  
The EIS line set covered the temperature range up to log~$T \sim$ 7.2 (Fe XXIV).  Emissions from the high-temperature lines (log~$T \ge 7.0$) are generally weak and blended with stronger, low temperature lines. 
Usually, these lines are only detected in flaring regions, and lines with temperatures above log~$T \sim 6.8$ are not customarily detected, as described in Section 2.2.
From 17:50 to 19:00 (including the \foxsi observation period), \hinode pointed to a quiescent region in the northwest of the solar disk in order to support \textit{FOXSI}'s quiet-Sun investigation.

\subsection{\hinode XRT}

\begin{figure*}[t]
 \begin{center}
  \includegraphics[width=16cm]{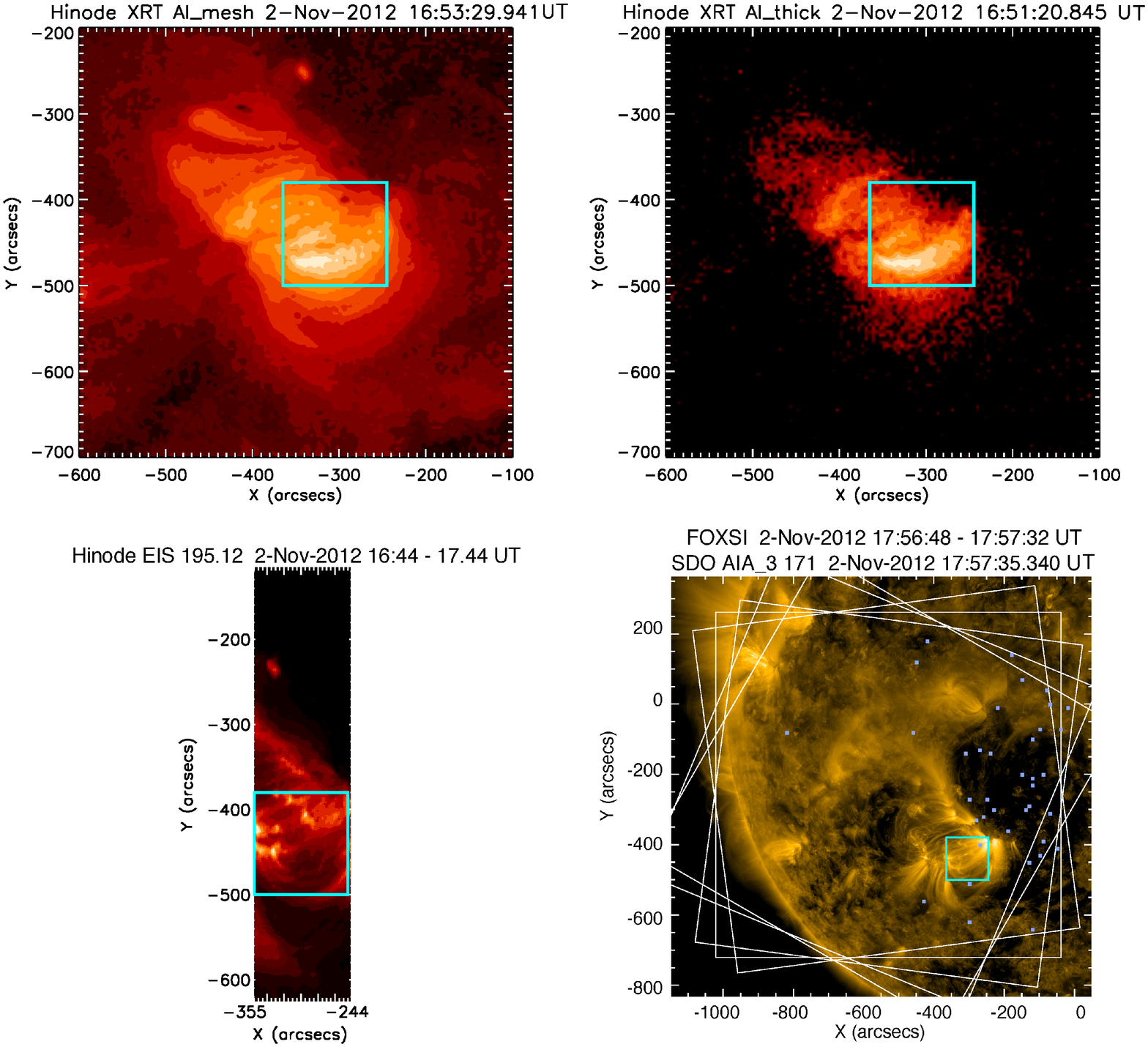} 
 \end{center}
\caption{Images of the active region from \textit{Hinode}/XRT, \textit{Hinode}/EIS,  \textit{SDO}/Atmospheric Imaging Assembly (AIA) and \textit{FOXSI}.  Top row: XRT images on log scales with the (left) Al mesh filter and (right) Al thick filter.  
Lower left: EIS intensity map of the Fe~XII 195.12~\AA ~line (log~$T \sim$6.1).  Lower right: \textit{SDO}/AIA 171~\AA ~image overlaid with the \foxsi observed counts from five detectors, collected over 44 seconds.  The field of view of each detector is shown by a white box.} \label{fig:image}
\end{figure*}

XRT observed AR 11602 with multiple filters. A set of images composed of 11 filter pairs were taken every 20 minutes
with a field of view (FOV) of 2048''$\times$2048'' and 4''$\times$4'' resolution (4$\times$4 CCD pixel binning).  
XRT images combining short and long exposures with the thin and thick filters (Al mesh and Al thick) are shown in the upper panels of Figure~\ref{fig:image}.  Significant emission is clearly detected with the Al thick filter. For the purposes of this paper the core of the active region is defined as the rectangular area shown in cyan in Figure~\ref{fig:image} (with corners [-365, -500] arcseconds and [-245, -380] arcseconds in helioprojective coordinates). This region is used for the analysis and discussion below.  
Using the standard SolarSoft procedure \verb+xrt_prep.pro+, 
we removed cosmic ray spikes and CCD biases, 
but we did not apply the Fourier clean and vignetting removal because they introduce uncertainties and the region of interest is almost at the center of the FOV.  
After applying the dark current model using
\verb+xrt_prep.pro+, we applied a further correction to set the zero-point at the median value of the lowest 5 CCD rows since these pixels are below the shielding material and not affected by the sky events.  
The averaged count rate for each filter pair in the region is shown in Table~\ref{tab:table_xrt}.  

\begin{table*}
  \caption{Observed active region XRT counts used for the DEM analysis.}
  \label{tab:table_xrt}
  \begin{center}
    \begin{tabular}{ccc}
      \hline 
			Filter pair		& Count rate   & Predicted count rate$^+$\\
        					& [DN s$^{-1}$ pixel$^{-1}$]$^{*}$  &  [DN s$^{-1}$ pixel$^{-1}$]$^{*}$   \\
      														\hline \hline 
 		Open / Be thick 	&   0.00663 	 &0.00728\\
    	Open / Al thick		&	0.173  &0.368\\
    	Open / Ti poly		&	207  &142\\
    	Open / Al mesh		&	498  &347\\
      	Al poly / Ti poly	&	114  &83.1\\
     	C poly / Ti poly	&	87.1  &68.2\\
      	C poly / Open  		&	279  &219\\
        Be thin / Open		&	11.5  &51.0\\
        Be med / Open		&	6.99  &8.93\\
        Al med / Open		&	3.80  &4.10\\
    	Al poly / Open		&	389  &297\\
      									\hline 
    \end{tabular}\\
    \small{(*) Per single pixel of the XRT CCD ($\sim$1''$\times$1''), not an area element of the 4$\times$4 binning.  (+) Predicted XRT count rates derived from the DEM solution.  }
  \end{center}
\end{table*}

\subsection{Hinode EIS}
\label{sec:eis}
EIS observed the active region using the 1-arcsecond-wide slit scanning over the region, for an FOV of 120$\times$512 arcsecond$^2$ for the complete raster.  EIS took spectra in 25 
wavelength windows with spatial sampling every 1 arcsec in the solar-Y direction (along the slit) and scanned every 2 arcsec in the solar-X direction.  
The slit scan was performed twice: once from 16:44 to 17:44~UT and again from 19:14 to 20:14~UT.
The observed intensity map of the Fe~XII line (log~$T \sim$6.1) for the first observation is shown in the lower left panel of Figure~\ref{fig:image}.

Values for all observed EIS lines are listed in Table~\ref{tab:table_eis}.  
The EIS data were calibrated using the pre-launch calibration \citep{lang2006}, and
then the \citet{warren2013b} time-dependent correction was applied. 
The values were obtained via Gaussian fits to the line profiles using spatially averaged spectra. 
Lines that are not clearly detected or are possibly blended are excluded from this table and from our analysis, except the Ca~XVII line.  
The Ca~XVII line was kept because it is expected to be sensitive to the highest temperatures of any line in this EIS set; its intensity is derived using the method of \citet{ko2009}.
In total, lines sensitive to temperatures from log~$T \sim$5.6 to 6.8  were detected.  
The plasma density can be estimated from the ratio of a pair of lines with an identical ion, such as the Fe~XII 186.88 / 195.12~\AA~ratio (\cite{young2007}).
From this ratio, the coronal plasma density in the active region is estimated to be 1.6$\times$10$^{9}$~cm$^{-3}$.

\begin{table*}
  \caption{Observed EIS line intensities at the active region used for the DEM analysis.  Shown line wavelengths and temperatures are from \cite{culhane2007}, \cite{young2007} and \cite{young2009}.}\label{tab:table_eis}
  \begin{center}
    \begin{tabular}{ccccc}
      \hline 
      Ion		&	Wavelength	&	Temperature	&	Intensity   & Predicted intensity$^+$ \\
       			&	[\AA ]		&	log~$T$ [K] &	[erg cm$^{-2}$ s$^{-1}$ str$^{-1}$] & [erg cm$^{-2}$ s$^{-1}$ str$^{-1}$] \\
      																\hline \hline 
		Fe VIII	&	185.21		&	5.6			&	217 &211 \\ 
		Fe VIII	&	186.60		&	5.6			&	118  &111\\
		Fe IX 	&	197.86		& 	5.9 			& 	47.7  &62.5\\
		Fe X 	& 	184.54		& 	6.0 			& 	374  &292\\
		Fe XI 	& 	180.40		& 	6.1 			& 	1474  &1445\\
		Fe XI 	& 	182.16		& 	6.1 			& 	303  &281\\
		Fe XII 	& 	186.88		& 	6.1 			& 	534  &439\\
		Fe XII 	& 	195.12		& 	6.1 			& 	1221  &1339\\
		Si X 	& 	258.37		& 	6.1 			& 	288  &364\\
		Fe XIII & 	201.13		& 	6.2 			& 	485  &362\\
		Fe XIII & 	202.04		& 	6.2 			& 	1072  &700\\
		S X 	& 	264.23		& 	6.2 			& 	80.6  &110\\   
		Fe XIV 	& 	264.79		& 	6.3 			& 	540  &708\\
		Fe XIV 	& 	270.52		& 	6.3 			& 	336  &363\\
		Fe XV 	& 	284.16		& 	6.3 			& 	4760  &4726\\
		S XIII 	& 	256.68		& 	6.4 			& 	293  &389\\
		Fe XVI 	& 	262.98		& 	6.4 			& 	232  &182\\  
		Ca XIV 	& 	193.87		& 	6.5			& 	24.3  &17.2\\
		Ca XVII & 	192.86		& 	6.8			& 	18.6  &3.31\\
      \hline 
    \end{tabular}\\
 \small{(+) Predicted EIS intensities derived from the DEM solution.  }
  \end{center}
\end{table*}

\subsection{\foxsi}
\label{sec:foxsi}

 \foxsi pairs seven direct-focusing optics modules with a dedicated silicon strip detector for each.  Each optics module is composed of seven Wolter-I mirrors produced using an electroformed nickel replication process at MSFC \citep{ramsey2002, ramsey2006}.  The detectors are double-sided silicon detectors designed at ISAS \citep{ishikawa2011dssd} and are based on the detector development for the Hard X-ray Telescope aboard the \textit{Astro-H} spacecraft \citep{kokubun2012}.  More detail on the \foxsi instrument and its first flight can be found in \citet{krucker2013foxsi} and \citet{krucker2014}.  Data are collected independently from the seven detectors, so each optic/detector pair provides an independent measurement.

During \textit{FOXSI}'s first flight, AR 11602 was observed during the first target pointing for 44 seconds.  The FOV is limited both by the off-axis response of the optics and the physical size of the detector.  The detector active area is 9.6$\times$9.6 mm$^2$, allowing a FOV of 16.5 arcminutes.  Because of imperfect alignment between the instrument and the rocket's pointing system, the on-axis position of the optics modules was located approximately 2 arcminutes from the pointing center of the rocket.  This offset was determined in-flight by comparing the microflare position as observed by \foxsi and by \textit{RHESSI}. 
The active region center was located $\sim$3 arcminutes from the rocket pointing center, or $\sim$5 arcminutes from the optics axis.  At 5 arcminutes off-axis, the optics efficiency is $\sim$85\% of the nominal on-axis value; this factor has been included in the instrument response.  

The optics modules were covered on both ends by thermal blankets composed of kapton and aluminized mylar to avoid thermal distortions of the mirror modules.  In addition to those nominal blankets, additional thermal blanketing material (located on the inner surface of the telescope metering structure) moved into and blocked the optical path sometime during the flight, reducing the effective area from its nominal value, mostly at low energies.  The effect of this additional blanketing is estimated by comparing observed \foxsi and \textit{RHESSI} thermal spectra from the microflare (which was observed by \foxsi during the second half of its observing time).  \rhessi measures a total emission measure of 4.8 $\pm1.0\times10^{46}$ cm$^{-3}$ and an electron temperature of 9.4 $\pm$0.4 MK for this microflare.  Based on the \textit{RHESSI} observed parameters, expected \textit{FOXSI} count spectra were computed and compared with the measured values, yielding the actual spectral response of the instrument in-flight.  
From this analysis, we estimate that \foxsi counts were reduced to 11\% and 24\% of their nominal values for the 6-7 and 7-8~keV bands by the excess blanketing.  These factors have been included in the instrument response for the results presented here.

It is also possible to perform a rough self-calibration of the additional blanketing absorption by deriving temperatures from ratios of \foxsi fluxes at different energies.  By requiring that the derived temperatures are constant across the observed energy range, we can fit the additional absorption for each detector.  Temperatures derived in this way agree with those derived from \rhessi data, confirming those results.  Details of this calibration will be shown in a following paper.
 
All seven detectors measured low count rates during the observation of AR 11602.  
The lower right panel of Figure~\ref{fig:image} shows the positions of all 4-15 keV photons measured by detectors 0, 2, 4, 5 and 6, a total of 38 counts.  Two of the seven detectors were excluded from this analysis.  Detector 1 showed excess noise on one half of the FOV due to a noisy readout ASIC; since this noise would interfere with the measurement of faint signals this detector was excluded.  Detector 3 was excluded because the corresponding optic shifted during launch; this was noted by visual inspection post-flight and is responsible for a blurring of the flare image.  
For this portion of the flight, one detector (\#4) measured a significantly higher count rate than the other detectors, leading us to believe that this detector was blocked by the least amount of additional blanketing.

\section{Analysis and Discussion}
\begin{figure}
 \begin{center}
  \includegraphics[width=8cm]{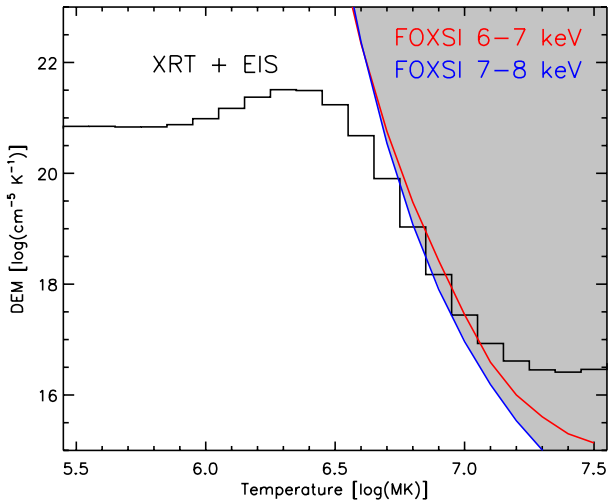} 
 \end{center}
\caption{\foxsi loci curve in two energy bands (red and blue) overplotted on a differential emission measure of the active region estimated by \textit{Hinode}/XRT and EIS (black).  The shaded region shows the parameter space that is excluded by the \foxsi observation.} \label{fig:dem}
\end{figure}

Figure~\ref{fig:dem} shows the DEM as estimated from combined \textit{Hinode} EIS and XRT observations of the active region, with \foxsi loci curves overlaid.

The combined \textit{Hinode} DEM was calculated using the CHIANTI database and inversion technique (\cite{monsignori1991}, 
the SSWIDL procedure \verb+chianti_dem.pro+).
This code is basically designed for line spectroscopy, and we modified the code to input the XRT count rates and temperature response functions.  
We assumed the measurement reliabilities of the EIS line intensities and XRT count rates are similar, 
and all the XRT and EIS measurements are equally weighted in the inversion.  
We applied the latest XRT calibration by \citet{narukage2014}, which provides a significantly upgraded XRT response function, especially for the thicker filters.  
The peak in the \textit{Hinode} DEM is located at log~$T \sim$ 6.3-6.4, suggesting the main coronal component of this active region is $\sim$2-2.5~MK plasma.  
The combined XRT and EIS DEM also shows the presence of an extremely hot component with a temperature $>$10~MK. 

The counts measured by \foxsi (shown spatially in Figure~\ref{fig:image}) do not come from the active region; rather, they are spread across the field of view.  These counts could be quiet-Sun flux or instrument artifacts (for example, single-bounce photons 
(`ghost rays') from the microflare on the west limb, which had already begun at the time of this pointing but which was well outside FOXSI's FOV at the time); this determination will be the focus of a separate paper.  In either case these counts constitute a background for the measurement of HXRs from the AR.  The number of counts in any given energy band are so small that precise background statistics cannot be determined.  We therefore assume that an AR HXR source would be detected if it produced a number of counts equal to the total number of counts in the image; this is a conservative estimate. 
Dividing the counts into 1 keV bins, we calculate the emission measure required for a variety of temperatures to produce the measured counts in each bin.  This produces a loci curve for the DEM for each 1 keV bin.  Curves in Figure~\ref{fig:dem} show these loci for the 6-7 and 7-8~keV bands.

We find that the \textit{Hinode}-derived hot ($>$8 MK) component in the DEM is inconsistent with the \foxsi observations in the gray area in Figure~\ref{fig:dem}.  If there is plasma with a DEM distribution extended into this region, \foxsi should have detected far more flux from the AR.  The hot component as derived mainly by XRT would have produced $>$1300~counts in the \foxsi spectrum above 6~keV, but none were detected.  
A similar result was found by \citet{schmelz2009b}. After including \rhessi HXR upper limits in a DEM analysis, they found that their originally reported high temperature component was greatly suppressed. \citet{winebarger2012} reports that the currently available EUV and SXR observatories have blind spots to high temperature plasma, at least in the presence of an intense component around a few MK as seen in non-flaring active regions. 
In addition, \citet{testa2012b} pointed out possible discrepancies between estimated DEMs from the EUV observations by comparing models and DEMs calculated by simulated \textit{Hinode}/EIS and \textit{SDO}/AIA data.  
The \foxsi observations highlight this issue and demonstrate the necessity of HXR observations to constrain the high-temperature end of the DEM.

The additional blanketing in the \foxsi instrument (see Section~\ref{sec:foxsi}) is impossible to perfectly calibrate because the degree of absorption could have changed in time due to further physical motions of the blankets.
Nevertheless, at high photon energies ($>10$ keV) where the blanket material contributes little absorption, the hot component derived mainly from XRT 
still predicts 
measured counts above 10 keV of a total of $>$200~counts, far more than were actually recorded. Hence, the blanketing issue alone cannot account for the discrepancies between \foxsi and \textit{Hinode}, and these discrepancies are not influenced by the cross-calibration with \textit{RHESSI}.

\section{Summary and Future Work}
We performed an active region DEM analysis using \textit{Hinode}/XRT and EIS, and compared it to \foxsi detection limits as indicated by the \foxsi loci curves.  
We find an inconsistency between the \textit{Hinode} DEM and the \foxsi limits for the hot, faint portion of the DEM with T$>$8~MK.
We conclude that the $>$8~MK component above 3$\times$10$^{44}$~cm$^{-3}$ is excluded from the DEM since the \foxsi detector performance 
is validated via the co-observation of a microflare with \textit{RHESSI}.

A second flight of \foxsi (\textit{FOXSI}-2) is scheduled for December 2014 and will include some upgrades.  \textit{FOXSI}-2 will provide an improvement to optics by including additional small-diameter shells.
In addition, we might update the detector by replacing some of the detectors with cadmium telluride double-sided strip detectors \citep{watanabe2009, ishikawa2010}. 
To avoid the issue of excess material in the optical path, inner thermal blankets will be removed, and so we therefore expect a combined increase in sensitivity relative \textit{FOXSI}-1 by about an order of magnitude. A future spacecraft based on the \foxsi technology would have the necessary sensitivity to systematically image hot components in non-flaring active regions, or set stringent upper limits to be used to distinguish different coronal heating models.


\section*{Acknowledgment}
\foxsi was funded by NASA's LCAS program, grant NNX11AB75G.  
We would like to thank Dr. Kyoung-Sun Lee for her help with the EIS DEM analysis. 
The work was supported through a Grant-in-Aid for JSPS Fellows from Japan Society for the Promotion of Science.  
Work at UC Berkeley received support from NASA GSRP fellowship NNX09AM40H.  
\textit{Hinode} is a Japanese mission developed and launched by ISAS/JAXA, collaborating with NAOJ as a domestic partner, NASA and STFC (UK) as international partners. Scientific operation of the \hinode mission is conducted by the \hinode science team organized at ISAS/JAXA. This team mainly consists of scientists from institutes in the partner countries. Support for the post-launch operation is provided by JAXA and NAOJ (Japan), STFC (U.K.), NASA, ESA, and NSC (Norway).



\begin{thebibliography}{}

\bibitem[Bryans et al.(2009)]{bryans2009} Bryans, P., Landi, E., 
\& Savin, D.~W.\ 2009, \apj, 691, 1540 

\bibitem[Cargill 
\& Klimchuk(2004)]{cargill2004} Cargill, P.~J., \& Klimchuk, J.~A.\ 2004, \apj, 605, 911 

\bibitem[Culhane et al.(2007)]{culhane2007} Culhane, J.~L., Harra, 
L.~K., James, A.~M., et al.\ 2007, \solphys, 243, 19 


\bibitem[Guennou et al.(2013)]{guennou2013} Guennou, C., Auch{\`e}re, F., Klimchuk, J.~A., Bocchialini, K., \& Parenti, S.\ 2013, \apj, 774, 31

\bibitem[Golub et al.(2004)]{golub2004} Golub, L., Deluca, E.~E., 
Sette, A., 
\& Weber, M.\ 2004, The Solar-B Mission and the Forefront of Solar Physics, 325, 217 

\bibitem[Golub et al.(2007)]{golub2007} Golub, L., Deluca, E., 
Austin, G., et al.\ 2007, \solphys, 243, 63 

\bibitem[Hannah \& Kontar(2012)]{hannah2012} Hannah, I.~G., \& Kontar, E.~P.\ 2012, \aap, 539, A146 

\bibitem[Ishikawa et al.(2010)]{ishikawa2010} Ishikawa, S., 
Watanabe, S., Fukuyama, T., et al.\ 2010, Japanese Journal of Applied 
Physics, 49, 116702 

\bibitem[Ishikawa et al.(2011)]{ishikawa2011dssd} Ishikawa, S., Saito, 
S., Tajima, H., et al.\ 2011, IEEE Transactions on Nuclear Science, 58, 
2039 

\bibitem[Ko et al.(2009)]{ko2009} Ko, Y.-K., Doschek, G.~A., 
Warren, H.~P., \& Young, P.~R.\ 2009, \apj, 697, 1956 

\bibitem[Kobayashi et al.(2011)]{kobayashi2011} Kobayashi, K., 
Cirtain, J., Golub, L., et al.\ 2011, \procspie, 8147

\bibitem[Kokubun et al.(2012)]{kokubun2012} Kokubun, M., Nakazawa, 
K., Enoto, T., et al.\ 2012, \procspie, 8443 

\bibitem[Kosugi et al.(2007)]{kosugi2007} Kosugi, T., Matsuzaki, 
K., Sakao, T., et al.\ 2007, \solphys, 243, 3 

\bibitem[Krucker et al.(1997)]{krucker1997} Krucker, S., Benz, 
A.~O., Bastian, T.~S., \& Acton, L.~W.\ 1997, \apj, 488, 499 

\bibitem[Krucker et al.(2013)]{krucker2013foxsi} Krucker, S., Christe, 
S., Glesener, L., et al.\ 2013, \procspie, 8862  

\bibitem[Krucker et al.(2014)]{krucker2014} Krucker, S., Christe, 
S., Glesener, L., et al.\ 2014, \apjl, \textit{submitted}

\bibitem[Landi et al.(2012)]{landi2012} Landi, E., Del Zanna, G., 
Young, P.~R., Dere, K.~P., \& Mason, H.~E.\ 2012, \apj, 744, 99 

\bibitem[Lang et al.(2006)]{lang2006} Lang, J., Kent, B.~J., 
Paustian, W., et al.\ 2006, \ao, 45, 8689 

\bibitem[Lin et al.(2002)]{lin2002} Lin, R.~P., Dennis, B.~R., 
Hurford, G.~J., et al.\ 2002, \solphys, 210, 3

\bibitem[Masuda et al.(1994)]{masuda1994} Masuda, S., Kosugi, T., 
Hara, H., Tsuneta, S., \& Ogawara, Y.\ 1994, \nat, 371, 495 

\bibitem[McTiernan(2009)]{mctiernan2009} McTiernan, J.~M.\ 2009, \apj, 697, 94

\bibitem[Miceli et 
al.(2012)]{miceli2012} Miceli, M., Reale, F., Gburek, S., et al.\ 2012, \aap, 544, A139 

\bibitem[Monsignori Fossi 
\& Landini(1991)]{monsignori1991} Monsignori Fossi, B.~C., \& Landini, M.\ 1991, Advances in Space Research, 11, 281 


\bibitem[Narukage et al.(2014)]{narukage2014} Narukage, N., Sakao, 
T., Kano, R., et al.\ 2014, \solphys, 289, 1029 

\bibitem[O'Dwyer et 
al.(2011)]{odwyer2011} O'Dwyer, B., Del Zanna, G., Mason, H.~E., et al.\ 2011, \aap, 525, A137 

\bibitem[Patsourakos 
\& Klimchuk(2009)]{patsourakos2009} Patsourakos, S., \& Klimchuk, J.~A.\ 2009, \apj, 696, 760 

\bibitem[Ramsey et al.(2002)]{ramsey2002} Ramsey, B.~D., 
Alexander, C.~D., Apple, J.~A., et al.\ 2002, \apj, 568, 432 

\bibitem[Ramsey(2006)]{ramsey2006} Ramsey, B.~D.\ 2006, Advances 
in Space Research, 38, 2985 

\bibitem[Reale 
\& Orlando(2008)]{reale2008} Reale, F., \& Orlando, S.\ 2008, \apj, 684, 715 

\bibitem[Reale et al.(2009)]{reale2009a} Reale, F., Testa, P., 
Klimchuk, J.~A., \& Parenti, S.\ 2009, \apj, 698, 756 

\bibitem[Reale et al.(2009)]{reale2009b} Reale, F., McTiernan, 
J.~M., \& Testa, P.\ 2009, \apjl, 704, L58 

\bibitem[Schmelz et al.(2009a)]{schmelz2009a} Schmelz, J.~T., Saar, 
S.~H., DeLuca, E.~E., et al.\ 2009, \apjl, 693, L131 

\bibitem[Schmelz et al.(2009b)]{schmelz2009b} Schmelz, J.~T., 
Kashyap, V.~L., Saar, S.~H., et al.\ 2009, \apj, 704, 863 

\bibitem[Schmelz et al.(2012)]{schmelz2012} Schmelz, J.~T., Kimble, 
J.~A., \& Saba, J.~L.~R.\ 2012, \apj, 757, 17 

\bibitem[Testa et al.(2011)]{testa2011} Testa, P., Reale, F., 
Landi, E., DeLuca, E.~E., \& Kashyap, V.\ 2011, \apj, 728, 30 

\bibitem[Testa 
\& Reale(2012)]{testa2012} Testa, P., \& Reale, F.\ 2012, \apjl, 750, L10 

\bibitem[Testa et al.(2012)]{testa2012b} Testa, P., De Pontieu, 
B., Mart{\'{\i}}nez-Sykora, J., Hansteen, V., 
\& Carlsson, M.\ 2012, \apj, 758, 54 

\bibitem[van Ballegooijen et al.(2011)]{vanballegooijen2011} van 
Ballegooijen, A.~A., Asgari-Targhi, M., Cranmer, S.~R., 
\& DeLuca, E.~E.\ 2011, \apj, 736, 3 

\bibitem[Warren et al.(2011)]{warren2011} Warren, H.~P., Brooks, 
D.~H., \& Winebarger, A.~R.\ 2011, \apj, 734, 90 

\bibitem[Warren et al.(2012)]{warren2012} Warren, H.~P., 
Winebarger, A.~R., \& Brooks, D.~H.\ 2012, \apj, 759, 141 

\bibitem[Warren et al.(2013)]{warren2013} Warren, H.~P., Mariska, 
J.~T., \& Doschek, G.~A.\ 2013, \apj, 770, 116 

\bibitem[Warren et al.(2013b)]{warren2013b} Warren, H.~P., 
Ugarte-Urra, I., \& Landi, E.\ 2013b, arXiv:1310.5324 

\bibitem[Watanabe et al.(2009)]{watanabe2009} Watanabe, S., 
Ishikawa, S.-N., Aono, H., et al.\ 2009, IEEE Transactions on Nuclear 
Science, 56, 777 

\bibitem[Winebarger et al.(2012)]{winebarger2012} Winebarger, A.~R., 
Warren, H.~P., Schmelz, J.~T., et al.\ 2012, \apjl, 746, L17 

\bibitem[Weber et al.(2004)]{weber2004} Weber, M.~A., Deluca, 
E.~E., Golub, L., 
\& Sette, A.~L.\ 2004, Multi-Wavelength Investigations of Solar Activity, 223, 321 

\bibitem[Young et al.(2007)]{young2007} Young, P.~R., Del Zanna, 
G., Mason, H.~E., et al.\ 2007, \pasj, 59, 857 

\bibitem[Young(2009)]{young2009} Young, P.~R.\ 2009, \apjl, 691, 
L77 

\end{thebibliography}
\end{document}